# On Determinism and Bell Test

Shoujiang Wang[1], Xiulan Wang[2*]

Abstract:

Many Bell test results violate Bell's inequality. The premise of Bell's inequality is local determinism. We propose that, it can't be proved that something's mechanism isn't deterministic; if loopholes are not the reason of violation of Bell's inequality, violation of Bell's inequality illustrates that the mechanism isn't local. We propose an explanation of Bell test, and the explanation illustrates the locality loophole is not closed by Weihs' experiment.

Many Bell tests [1,2,3] have been performed. The loophole related to detection has been closed in several independent experiments [4,5,6], and their results violate Bell's inequality. Aspect's experiment in 1982 [7] and Weihs' experiment in 1998 [8] attempt to close the locality loophole, and their results also violate Bell's inequality. The premise of Bell's inequality is local determinism. So people may think that the experimental results don't have a local deterministic explanation. In this paper, one cognitive deficit of people is proposed to illustrate that it can't be proved that the mechanism isn't deterministic. If loopholes are not the reason of violation of Bell's inequality, violation of Bell's inequality illustrates that the mechanism isn't local. We propose an explanation of Bell test. The explanation illustrates Weihs' experiment in 1998 [8] does not close locality loophole.

[1] School of Chemical Engineering and Technology, Xi'an Jiaotong University, Xi'an, 710049, China.
 [2]School of Electronic Information and Electrical Engineering, Shanghai Jiao Tong University, Shanghai, 200030, China.
[*] Corresponding author. E-mail: axnhwxl@163.com

On determinism

It can't be proved that the mechanism isn't deterministic. We consider that the meaning of determinism (or causality) is that different results must have different causes, so causes decide results. The meaning of indeterminism is opposite to the meaning of determinism, so the meaning of indeterminism is that the different results may have the same causes. If the real mechanism is deterministic, it can't be proved that the mechanism isn't deterministic. If the real mechanism is nondeterministic, then suppose that every detail in the experiment is reviewed, first, if different results have different causes in all details, it can't be proved that the mechanism isn't deterministic; second, if some details appear different results under the same causes, then we can consider that the states are different in the basis of real causes (if real causes include state, we can consider that the state in real causes describes the part of the state, and the other part of state is different), then it can also be considered that different results have different causes in these details, and the real mechanism in the other part is adopted, and then a deterministic explanation of the experimental results is formed, this explanation may be difficult to be acquired, but it exists. Although some inexistent causes are adopted, people cannot determine whether the causes which are actually fictive are inexistent causes or existent causes which have not been observed, it is difficult to illustrate that the deterministic explanation of the experimental results is wrong, then it can't be proved that the mechanism isn't deterministic. In the following, we will explain the questions that may be proposed. Suppose that someone question whether the unreal explanation of second situation of indeterminism is denied by axiom, for example, suppose that the explanation appears superluminal transmission of information, and suppose that there is an axiom that the transmission speed of information can't exceed the speed of light, because the explanation is real mechanism except the added state which can't be illustrated inexistent, it can only be said that the axiom is wrong. If people think that a factor has an effect on the result through the experiment result, but there is not the factor in the explanation, we can explain that the factor's influence on the results is a coincidence, and then it can't be sure that the explanation is wrong. But in reality, people do use the logic which is not strict, and they may consider that it has been proved that the explanation is wrong in a non-strict way. There is not the factor in the real effects if there is not the factor in the explanation according to our construction method. So it should not be proved that the explanation is wrong even in a non-strict logic which is useful and used by people. If the real mechanism is nondeterministic, and if someone proves that the mechanism isn't deterministic and if there isn't mistake in proof, a possible reason may be the use of the knowledge which cannot be strictly confirmed. As can be seen from the preceding analysis, even if the real mechanism is not deterministic, we cannot find the experimental evidence which can make us be partial to the indeterminism, so some knowledge used in proof not only may not be strictly confirmed, but also may not be confirmed in a non-strict way and so the proof can't enhance the credibility of the conclusion. And in fact we cannot be sure that the real mechanism is nondeterministic. If the real mechanism is deterministic, the viewpoint that it can be proved that the mechanism is nondeterministic may impede the people to explore the true mechanism, so we'd better not recognize the viewpoint that it can be proved that the mechanism is nondeterministic.

On Bell tests

Suppose that some experimental results do not exist a local deterministic explanation, first, it can be certain that the real mechanism is not local deterministic, second, if the real mechanism is local nondeterministic, there must be real local explanation; if different results have different causes in this real local explanation, it is contradictory to the precondition that local deterministic explanation does not exist; if some different results have the same causes in this real local explanation, the local nondeterministic explanation can be turned into local deterministic explanation by artificially supposing that the states are different, it is contradictory to the precondition, so the real mechanism is also not local nondeterministic. That is to say that if the existence of the local deterministic explanation is denied, it needs to be admitted that the real mechanism is non-local because the real mechanism is not local deterministic or local nondeterministic.

Einstein, Podolsky, and Rosen show that quantum mechanics could not be a complete theory in the premise of realism [9]. In Bell's paper, it is shown that when the theory becomes causality and locality by additional variables, the theory will be incompatible with the statistical predictions of quantum mechanics [1]. So the premise of Bell's inequality is local determinism. Or to be more exact, existing a local deterministic explanation is the premise of Bell's inequality. But some literatures use the term 'local realism' as the premise of Bell's inequality [2,3]. Some people might think that under local condition realism can prove determinism. As can be seen from the preceding analysis, if the real mechanism is local, there must be a local deterministic explanation, so it is unnecessary to prove that the real mechanism is deterministic. And in fact we need not consider the thing about realism.

If all loopholes are closed or are not the reason of violation of Bell's inequality, there is not a local deterministic explanation of Bell test because the premise of Bell's inequality is local determinism. As can be seen from the preceding analysis, it illustrates that the real mechanism is non-local and it can't illustrate that the mechanism isn't deterministic.

Until now, there is not a Bell test which closes all loopholes. It seems to be accepted by the community that the experiment by Weihs [8] closes the locality loophole, but actually it does not close the locality loophole. In the experiment, the directions of polarization analysis were switched after the photons left the source. The experimental results violate Bell's inequality, but there may be the following loopholes:

(1) The experimental errors may make the results violate Bell's inequality coincidently. But this possibility is very small.

(2) In Weihs' paper [8], they said: "Selection of an analyzer direction has to be completely unpredictable, which necessitates a physical random number generator. A pseudo-random-number generator cannot be used, since its state at any time is predetermined." But if the mechanism of physical random number generator is deterministic, the random is pseudorandom, and as can be seen from the preceding analysis it can't be proved that the mechanism isn't deterministic, so it cannot be sure that selection of an analyzer direction is completely unpredictable. In a strict logic, their experiment cannot close the locality loophole because prediction cannot be excluded. But people do use the logic which is not strict and we think that the non-strict logic should be used, and we think a pseudo-random-number generator can be used even though its state at any time is predetermined. In addition, there are other ways except prediction to make use of pseudorandom, for example physical random number generator and the emission of photon pairs are controlled to violate Bell's inequality by something.

(3) Because of detection efficiency, only few photon pairs are detected. Though the loophole related to detection has been closed in several independent experiments [4,5,6], these experiments only show that there is other reason to make the results violate Bell's inequality and they cannot make sure that detection efficiency has not effect. But these experiments enhance the credibility of Bell's opinion [1]: "it is hard for me to believe that quantum mechanics works so nicely for inefficient practical set-ups and is yet going to fail badly when sufficient refinements are made."

(4) Bell's inequality is a law obtained by the average of all the states. The photon pairs which are emitted maybe cannot represent all the states. But this possibility is very small.

Although there are loopholes mentioned above, but we do not believe that these loopholes result in violation of Bell's inequality. If Einstein's theory of relativity is right, there should be other loophole. We propose an explanation of Bell test, the explanation illustrates the locality loophole is not closed by Weihs' experiment.

An explanation of Bell test

The Bell's theorem requires that the measurements of both ends are independent, but we think the measurements of both ends actually affect each other. The information of one end is transmitted to the other end in a speed which is not superluminal. The information is not transmitted before the measured object contacts the thing which has the information. We conjecture that the information is transmitted through the measured object.

Because photon transmits very fast, people may think that there is not enough time to transmit information if the Bell test uses photon as the measured object. So we illustrate in detail the Bell test which uses photon as the measured object.

When a photon contacts a measurement device, it does not have significant effect on the measurement device instantaneous, but it may have no effect or slight effect on the measurement device during a period of time. We propose a measurement mechanism of photon, this mechanism may not be real, but it can figuratively explain our idea: in the non-vacuum condition, single photon is not a point in the transmission process, and when the photon contacts the measurement device, it has slight effect on the measurement device because the energy is not concentrated. During a period of time, the photon is absorbed competitively at multiple locations, and then the photon has significant effect on a location of the measurement device.

In the closed interferometer configuration of Jacques's experiment [10], a single-photon pulse is split and travels through two paths and then is recombined to realize interference. Many experiments with different path were done. We think that the two parts of most photons in the experiments cannot reach the interference location at the same time through two paths, that is to say that some information may arrive ahead before the occurrence of the interference. While the final measurement result is determined by the interference, if a photon has significant effect on the measurement device instantaneous when it contacts the measurement device, it is not easy to explain that the information before the occurrence of the interference has no effect on the measurement in all cases of the experiments.

In Bell test, when the photon in entangled state contacts the measurement device, it has not significant effect on the measurement device during a period of time which may be different from the time of common photon and is enough to transmit the information.

The explanation illustrates Weihs' experiment in 1998 [8] does not close locality loophole. Fig.1 a is the original spacetime diagram of their paper [8]. They think that a photon has significant effect on the measurement device instantaneous when it contacts the measurement device at spacetime points "Y" and "Z". But according to our preceding analysis about Jacques's experiment [10], it is wrong. Fig.1 b is the modified spacetime diagram according to our explanation. The photons contact the measurement devices at spacetime points "Y" and "Z", respectively. But the photons have significant effect on the measurement devices at spacetime points "Y1" and "Z1", respectively. So the measurement can affect each other. If Einstein's theory of relativity is right, there must be loophole in Bell test. So we think our explanation is likely to be correct. Compared with common photon, the photon in entangled state may have different time during which the photon has not significant effect on the measurement device. For example, supposing photon is not a point in the transmission process, the shape of the photon may be affected by each other, and then the time may be affected. In our opinion, Jacques's experiment [10] may illustrate that a photon is not a point in the transmission process.

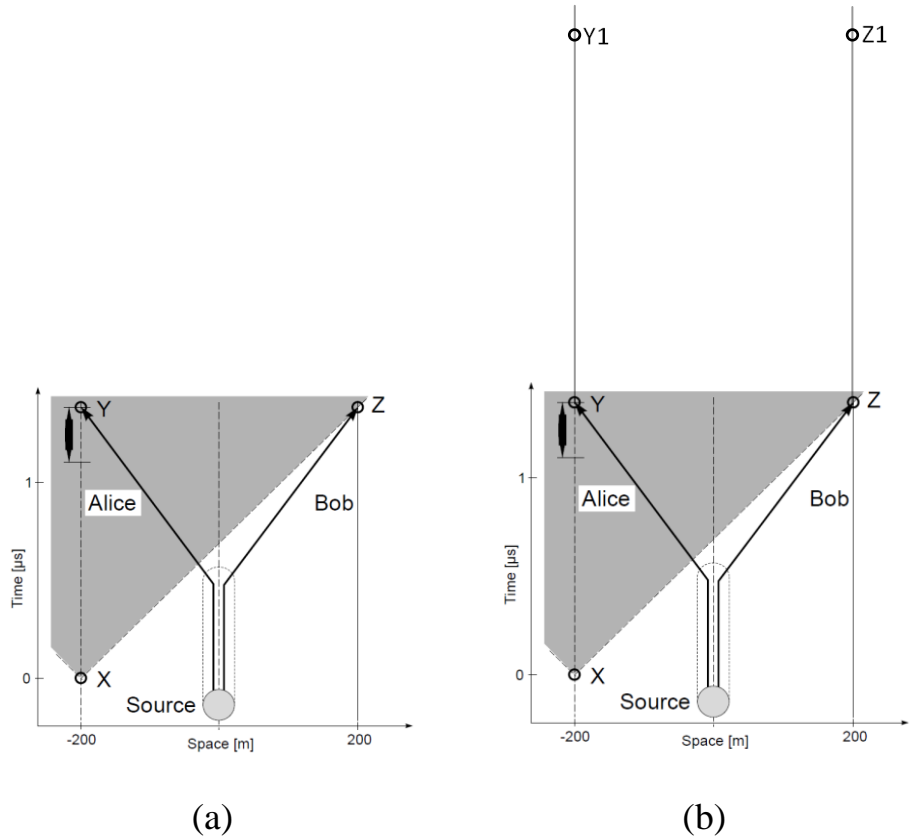

(a) (b)

FIG. 1. (a) Original spacetime diagram of Weihs' Bell experiment. (b) Modified spacetime diagram according to our explanation.

The new explanation of Bell test can be judged by Weihs' experiment. Fig.2 is the modified figure of Weihs'paper [8]. We only add three points (A, B and C) in the original figure of their paper. In the experiment, the measurement result and the analyzer setting are recorded with time [8]. Because the analyzer setting is changed very fast, we think they must adjust the time delay to make the analyzer setting correspond to the measurement result. Then you can indirectly estimate the measuring time (or the lower limit of the measuring time) by the time delay of the other parts which compose loop with the measurement device. Suppose that $t_1$, $t_2$ and $t_3$ is the time to translate message from A to B, B to C and A to C respectively, then $(t_3-t_1)$ is the estimated value of $t_2$. If the new explanation of Bell test is right, the indirect estimation of the measuring time is larger than the value which people usually estimate. But some time delay is not published in their paper. So an experiment similar to Weihs' experiment is worth doing.

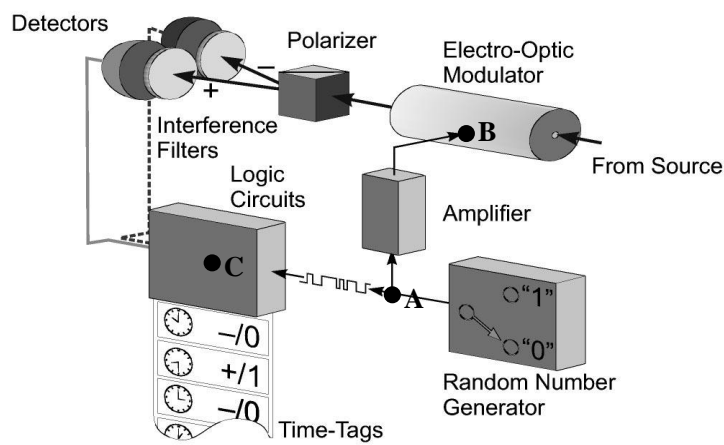

FIG.2. One of the two observer stations of Weihs' experiment.

Conclusion

Even if the phenomena of the same causes leading to different results are indeed observed, people still cannot determine whether there are causes which have not been observed and cause the results different. So people cannot find the evidence which makes people be partial to the indeterminism. So, as long as you choose to believe Albert Einstein's conception of "I cannot believe that God plays dice with the Cosmos", no one can illustrate that your opinion is wrong. If you choose to believe the indeterminism, although this opinion cannot be strictly proved to be wrong, because even if we find that the same reasons always have the same results, it can still be explained as coincidences, but the evidence is undoubtedly partial to the determinism and in fact denies the indeterminism in a non-strict way. Our conclusion also rejects all attempts to prove indeterminism. Quantum mechanics is not a deterministic theory, so people can never prove that quantum mechanics is a complete theory. If loopholes are not the reason of violation of Bell's inequality, violation of Bell's inequality illustrates that the mechanism isn't local. We propose an explanation of Bell test. If the explanation is proven to be correct by experiment, it will lead to a new awareness of photon. And we conjecture that it is the reason of violation of Bell's inequality that the measured objects in entangled state are not separated. Can the measured objects in entangled state be separated spontaneously?

Acknowledgements: The authors thank Dr. Weihs for replying our letters. The analysis about his experiment in 1998 does not represent the views of Dr. Weihs.